\input amstex
\documentstyle{amsppt}
\input boxedeps
\SetRokickiEPSFSpecial
\loadbold
\define\bcdot{\boldsymbol\cdot}
\define \iso{\overset \sim \to \to }
\define \Z{\Bbb Z}
\define \zt{\Z_2}

\define \im{\operatorname{im}} 
\define \Ainf{A_{\infty}}
\define \CA#1{C^{#1}(A)}
\define \CCA#1{CC^{#1}(A)}
\define \HA#1{HH^{#1}(A)}
\define \Hm{\operatorname{Hom}}
\define \Hom#1#2{\Hm(A^{#1},{#2})}
\define \fa#1#2{f(a_{#1} ,\dots , a_{#2})}
\define \fat#1{\widetilde f(a_1,\dots ,a_{#1})}
\define \fai#1{f(a_1,\dots ,a_ia_{i+1},\dots ,a_{#1})}
\define \tns{\otimes}
\define \mtns#1#2{{#1}_1\otimes \dots \otimes {#1}_{#2}}
\define \tmu{\widetilde \mu}
\define \dd{\widehat{D}}
\define \dt{\widetilde{D}}
\define \sg{\sigma_\Gamma}
\define \R{\Bbb R}
\define \Rgn{{\Cal R}_{g,n}^{met}}
\define \Mgn{{\Cal M}_{g,n}}
\define \Gor{\Gamma_{\text{\rm or}}}
\define \Gc{\Gamma_{\text{\rm cil}}}
\define \tm{\widetilde m}

\define \tepr{\otimes}

\topmatter
\title $A_{\infty}$ Algebras and the Cohomology of Moduli Spaces
\endtitle
\author
Michael Penkava 
\thanks The work of the first author was partially supported by NSF
Grant DMS-9404111.
\endthanks
and Albert Schwarz
\thanks  The work of the second author was partially supported by NSF Grant
 DMS-9201366. Research at MSRI is supported by NSF Grant
DMS-9022140.
\endthanks
\endauthor
\address
Department of Mathematics, University of California,
Davis, CA 95616
\endaddress
\email michae\@math.ucdavis.edu, schwarz\@math.ucdavis.edu
\endemail
\endtopmatter

\document      

\head \S 1. Introduction \endhead
Let us consider an $\Ainf$ algebra with an invariant inner product.
The main goal of this paper is to classify the infinitesimal 
deformations of this $\Ainf$ algebra preserving the inner product
and to apply this result to the construction of homology classes on
the moduli spaces of algebraic curves. With this aim, we define
cyclic cohomology of an $\Ainf$ algebra and show that it classifies 
the deformations we are interested in. To make the reading of our
paper more independent of other works, we include a short review
of Hochschild and cyclic cohomology of associative algebras, and
explain the definition of $\Ainf$ algebras.
\endgraf
Our constructions are based on ideas and results of Maxim Kontsevich;
moreover, he has informed us that he also has given a definition of
the cyclic cohomology of $\Ainf$ algebras in a different manner than
we do and has proved the results mentioned above as well. Another
definition of cyclic cohomology of $\Ainf$ algebras was given by
Ezra Getzler and John D.S. Jones \cite{5}. 
We did not study its relation to our definition.
\endgraf
In this paper we make the notational convention
when dealing with $\zt $ grading, that if $a$ is a homogeneous element 
with parity $|a|$, in superscripts we will use $a$ in place of $|a|$,
so that $ (-1)^a $ stands for $ (-1)^{|a|} $, and similarly, 
$ (-1)^{ab}=(-1)^{|a||b|} $, not $ (-1)^{|ab|} $, which is of course given
by $ (-1)^{a+b} $.
\head \S 2. Cohomology of Associative Algebras \endhead
In this section, we recall the definition of Hochschild cohomology
of associative algebras. We relate this notion to the 
theory of deformations of the associative algebra structure. Then we
discuss the theory of deformations of an associative algebra preserving
an invariant inner product, and relate this notion to cyclic
cohomology. The purpose of this is to motivate the definition of 
$\Ainf $ algebras, and to set the stage for the more general discussion 
of cohomology of $\Ainf $ algebras.
\endgraf
In this section, suppose that $A$ is an algebra over a field $k$. 
For 
simplicity, we suppose that $A$ is finite-dimensional over $k$.  Let 
$ \CA n =\Hom nA $ be the space of $n$-multilinear functions on $A$
with values in $A$; we call $\CA n$ the module of cochains of degree 
$n$ on $A$ with values in $A$. We define a coboundary operator
$b:\CA n \to \CA {n+1} $ by
$$
\aligned
b\fa 1{n+1}=&a_1\fa 2{n+1}\\
	    &+\sum_{i=1}^n(-1)^if(a_1,\dots ,a_ia_{i+1},\dots ,a_{n+1})\\
	    &+(-1)^{n+1}\fa 1{n+1}
\endaligned
\tag 1
$$
Then 
$$
\HA n=\ker (b:\CA n \to \CA {n+1})/\im(b:\CA {n-1} \to \CA n)
$$
is the Hochschild cohomology of $A$ with coefficients in $A$.
In this paper we do not consider cohomology with other coefficients. 
The connection between Hochschild cohomology and (infinitesimal) 
deformations of $A$ is given by $ \HA 2 $. If we denote the product in 
$A$ by $m$, an infinitesimally deformed product by $m_t$, and express 
$ m_t=m+t\phi $, where $ t^2=0 $, then the map $ \varphi:A^2 \to A $ is 
a Hochschild cocycle and the trivial deformations are given by Hochschild 
coboundaries. To show the first assertion note that 
$$
\align
m_t(m_t(a_1,a_2),a_3))=&a_1a_2a_3+t(\varphi(a_1,a_2)a_3+\varphi(a_1a_2,a_3)),
\\
m_t(a_1,m_t(a_2,a_3))=&a_1a_2a_3+t(a_1\varphi(a_2,a_3)+\varphi(a_1,a_2a_3)).
\endalign
$$
Associativity of $m_t$ is equivalent to the condition
$$
a_1\varphi(a_2,a_3)-\varphi(a_1a_2,a_3)+\varphi(a_1,a_2a_3)-
\varphi(a_1,a_2)a_3=0.
\tag 2
$$
But this last condition is simply the condition $ b\varphi=0 $.
\endgraf
On the other hand, the notion of a trivial deformation is given
by the condition that $A$ with the new multiplication is isomorphic
to the original algebra structure. This means that there is a linear 
bijection $ \rho_t:A\to A $ such that $ m_t(\rho_t(a_1),\rho_t(a_2))
=\rho_t(a_1a_2)$. We can express 
$ \rho_t=I+t\lambda $, where $ \lambda :A \to A $ is a linear map.
Then
$$
\align
m_t(a_1,a_2)&=\rho_t(\rho_t^{-1}(a_1),\rho_t^{-1}(a_2))\\
&=a_1a_2-t(a_1\lambda(a_2)-\lambda(a_1a_2)+\lambda(a_1)a_2)=
a_1a_2-t(b\lambda)(a_1,a_2).
\endalign
$$
Thus coboundaries give rise to trivial deformations. 
\endgraf
\define\inpr#1#2{\langle{#1},{#2}\rangle}
Next, consider an invariant non-degenerate inner product on $A$, 
by which we mean an inner product $ \inpr{\bcdot}{\bcdot} $ that satisfies 
$\inpr {ab}c=\inpr a{bc} $.
Note that for an invariant inner product, we also have that
$ \inpr {ab}c=\inpr {ca}b $, so that it is invariant under the cyclic 
permutations of $a ,b, c$.
We consider deformations of $A$ preserving this inner product, and these 
are governed by cyclic cohomology. To see this connection we first note 
that in the presence of an inner product, there is a natural isomorphism 
$ \Hom n{A}\iso \Hom {n+1}k $,
given as follows.
If we denote the image of $ f\in \Hom nA $ in $ \Hom {n+1}k $
by $ \widetilde f$, then
$$
\fat {n+1}=\inpr {\fa 1n}{a_{n+1}}.
$$
If we define an element $\widetilde f$ in $ \Hom nk $ to be cyclic whenever
$$
\fat n=(-1)^{n+1}\widetilde f(a_n,a_1,\dots ,a_{n-1}),
$$
then we see that a cyclic element $ \widetilde\varphi $ in $ \Hom 3k $ 
corresponds to a deformation $ \varphi $ in $ \CA 2 $ preserving the inner 
product, because
$$
\inpr {\varphi (a,b)}c=\widetilde\varphi (a,b,c)=
\widetilde\varphi (b,c,a)=\inpr a{\varphi (b,c)}
\tag 3
$$
\endgraf
A trivial deformation preserving the inner product is determined by
a linear map $ \rho_t=I+t\lambda $ as before, but in addition we assume 
that 
$$
\langle \rho_t(a_1),\rho_t(a_2)\rangle =\inpr {a_1}{a_2}.
\tag 4 
$$
This is equivalent to the condition
$ \inpr {\lambda(a_1)}{a_2}=-\inpr {a_1}{\lambda(a_2)} $. 
This latter condition is precisely the condition that 
$ \widetilde\lambda(a_1,a_2)=-\widetilde\lambda(a_2,a_1) $. 
In other words, $ \lambda $ is cyclic.
\endgraf
The  Hochschild  coboundary  operator  $b$  induces  a  coboundary  operator
  $\quad \widetilde b:\Hom {n+1}k \to \Hom {n+2}k $
by $ \widetilde b\widetilde f=\widetilde{bf} $.
As we shall show later, the coboundary operator takes cyclic elements to 
cyclic elements. If we denote the submodule of $ \Hom {n+1}k $ consisting of 
cyclic elements by $\CCA n $, then the cyclic cohomology of $A$ is defined by 
$$
H\CA n=
\ker(\widetilde b:\CCA n\to \CCA {n+1})/\im(\widetilde b:\CCA {n-1}\to
\CCA n) 
$$
\endgraf
This definition depends on the choice of the inner product. However, 
one can express the coboundary operator $\widetilde b$ without
reference to the inner product. If $f\in \Hom {n+1}k $, then we see that
$$
\aligned
\widetilde b &\fat {n+2}=\widetilde{bf}(a_1,\dots ,a_{n+2})=
\inpr {b\fa 1{n+1}}{a_{n+2}} \\
&=\inpr {a_1\fa 2{n+1}}{a_{n+2}} 
+\sum_{i=1}^n(-1)^i\inpr {\fai {n+1}}{a_{n+2}} \\
&\quad +(-1)^{n+1}\inpr {\fa 1na_{n+1}}{a_{n+2}} \\
&=\inpr {\fa 2{n+1}}{a_{n+2}a_1} 
+\sum_{i=1}^n(-1)^i\inpr{\fai {n+1}}{a_{n+2}} \\
&\quad +(-1)^{n+1}\inpr {\fa 1n}{a_{n+1}a_{n+2}} \\
&=\widetilde f(a_2,\dots ,a_{n+1},a_{n+2}a_1) 
+\sum_{i=1}^{n+1}(-1)^i\widetilde f(a_1,\dots ,a_ia_{i+1},\dots ,a_{n+2})
\endaligned
\tag 5
$$
If $\widetilde f$ is cyclic, then we see that 
$$
\aligned
\widetilde b &\fat {n+2}\\
&\qquad=(-1)^n\widetilde f(a_{n+2}a_1,a_2,\dots ,a_n)
+\sum_{i=1}^{n+1}(-1)^{ni+n+1}\widetilde f(a_ia_{i+1},\dots ,a_{i-1})\\
&\qquad=\sum_{i=0}^{n+1}(-1)^{n[i]+n+1}
\widetilde f(a_{[i]}a_{[i+1]},\dots ,a_{[i-1]}),
\endaligned
\tag 6     
$$
where $ [i]=i\pmod{n+2} $, and we make the convention that $ a_0=a_{n+2} $.
The fact that $ \widetilde b\widetilde f $ is cyclic follows easily from this 
formula for the differential.
\endgraf
Thus the cyclic cohomology characterizes the deformations of $A$ that
preserve an invariant inner product, independently of the particular
inner product involved. We note that cyclic homology is usually based on 
applying the operator $b$ to the complex 
$ C ^n(A,A^*)=\Hom n{A^*} $. However, 
the description we give here is equivalent to this one, 
because the inner product
induces a natural isomorphism between $ \Hom n{A^*} $ and 
$ \Hom nA $. Of course, the description in terms of 
$ C^n(A,A^*) $, being independent of any inner product,
makes completely transparent the fact that the cyclic homology does not
depend on the inner product. However, it does not elucidate the connection
between  $HC^2(A)$  and deformations preserving an inner 
product. As was pointed out to us by Getzler, this relation was
first shown by Connes-Flato-Sternheimer in \cite{1}. 

\head \S 3. Cohomology of $\zt $-graded algebras\endhead

Let us consider
a $\zt $-graded algebra $A$ over a ring $k$. The notion of an invariant inner
product needs to be modified in this case.
The definition of a graded symmetric inner product 
is given by the formula 
$$
\inpr ab=(-1)^{ab}\inpr ba
$$
for homogeneous elements $a$, $b$ in $A$. 
Let $k$ be a commutative ring. 
Consider elements of $k$ as having degree 0, making $k$
a $\zt $-graded ring in a trivial sense. If we require the inner 
product to be an even map when considered as a map from 
$ A\tepr A $ to $k$, then $\inpr ab=0 $ unless $a$ and $b$ have the
same parity. On elements of even parity the inner product is symmetric,
and on elements of odd parity the inner product is a skew symmetric
form. This definition is for an even inner product.
One can also consider the case when $k$ is a supercommutative ring.
Odd inner products can also be defined, but we will not consider them
in this paper. The notion of an invariant inner product is given by the 
same relation $\inpr {ab}c=\inpr a{bc} $, but now we have
$$
\aligned
  \inpr {ab}c&=\inpr a{bc}=(-1)^{(b+c)a}\inpr{bc}a=(-1)^{(b+c)a}\inpr b{ca}
\\
&=(-1)^{(b+c)a+(c+a)b}\inpr{ca}b=(-1)^{c(a+b)}\inpr{ca}b
\endaligned
\tag 7
$$
\endgraf
We also assume that the multiplication is an even map, so that 
$ |ab|=|a|+|b| $. We are only interested in deformations of 
$A$ that preserve this property. If we denote the deformed
multiplication by $ m_t=m+t\varphi $, and consider $t$ as an even
parameter, then $\varphi $ must be even map, but if $t$ is an odd
parameter, then $\varphi $ must be odd. For parameters, we assume the
property of graded commutativity, so that $ ta=(-1)^{ta}at $.
The associativity condition (2) is modified by this consideration,
so that now we have the formula
$$
(-1)^{a_1\varphi } a_1\varphi (a_2,a_3)-\varphi (a_1a_2,a_3)+
\varphi (a_1,a_2a_3)-\varphi (a_1,a_2)a_3=0.
\tag 8
$$
which is the deformation condition  for any homogeneous bilinear map
$\varphi $.
\endgraf
Now we consider a trivial deformation, which, as before, is given by
a linear map $ \rho_t=I+t\lambda :A\to A $, but the 
parameter is allowed to be odd, in which case $\lambda$ is odd as well. In this
case, a trivial deformation is given by
$$
m_t(a_1,a_2)=a_1a_2-t(a_1\lambda(a_2)-\lambda(a_1a_2)
+(-1)^{a_1}a_1\lambda(a_2)).
$$
These two results suggest that we should define
the Hochschild coboundary operator in the $\zt $-graded
case as the map $ b:\CA n\to \CA {n+1} $ given by
$$
\aligned
b\fa 1{n+1}={}&(-1)^{a_1f}a_1\fa 2{n+1} \\
&+\sum_{i=1}^{n}(-1)^i\fai {n+1}\\
& +(-1)^{n+1}\fa 1na_{n+1}
\endaligned
\tag 9
$$
for homogeneous $f$. It is easily checked that $ b^2=0 $.
Then deformations of $A$ correspond to cocycles,
while trivial deformations correspond to coboundaries of this new
Hochschild coboundary operator.

The definition of a cyclic element of $\CCA n$ is adjusted to be 
consistent with the grading, which makes the notion compatible with
the definition of the invariance of the inner product. 
This is accomplished by defining an element $ \widetilde f\in \CCA n $ 
to be cyclic if and only if
$$
\fat {n+1}=(-1)^{n+a_{n+1}(a_1+\dots + a_n)}
\widetilde f(a_{n+1},a_1,\dots ,a_n).
\tag 10
$$
Note that if $ \tm \in \CCA 2$ is given by $ \tm (a,b,c)=
\inpr {ab}c $, then $ \tm $ is cyclic precisely when the inner product 
is invariant with respect to the multiplication.
Also, we see that $\widetilde f$ is cyclic precisely when
$$ 
\inpr {\fa 1n}{a_{n+1}}=(-1)^{n+a_1f}\inpr {a_1}{\fa 2{n+1}} 
\tag 11
$$
for all homogeneous elements.
\endgraf
We want to express the Hochschild differential of cyclic homology
in terms of this new notion of cyclicity. 
Let us restrict ourselves to the case when $k$ is a field.
Then, as in the nongraded case,
there is a natural isomorphism between $\Hom n{A} $ and $\Hom {n+1}k $
defined in the same manner as before, inducing a coboundary operator
$\widetilde b:\Hom nk\to \Hom {n+1}{A} $ 
given by $ \widetilde b\widetilde f=
\widetilde{bf} $.
The proof of the formula below is straightforward, and cyclicity is 
easily seen from this formula.
\proclaim{Lemma}
The Hochschild differential $\widetilde b$ takes cyclic elements
to cyclic elements. For a cyclic element $\widetilde f$, the Hochschild
differential can be expressed in the form 
$$
\aligned
&\widetilde b\fat {n+1} \\
&=\sum_{i=0}^n
{-1}^{ni+n+1+(a_1+\dots +a_{[i-1]})(a_{[i]}+\dots +a_{n+1})}
\widetilde f(a_{[i]}a_{[i+1]},a_{[i+2]},\dots ,a_{[i-1]}),
\endaligned
\tag 12
$$
where $ [i]=i\pmod{n+1} $ and by convention, $ a_0=a_{n+1} $.
\endproclaim
The sign in the expression above, as in the formulas below, follow
from the exchange rule which is given by the principle that when two
elements are exchanged in an expression, the corresponding sign is $-1$ 
to the power equal to the product of the parities of the two elements.
\endgraf
As in the nongraded case, one sees that the cyclic cohomology $HC^2(A)$
classifies the deformations of $A$ which preserve a graded
symmetric inner product. Cyclic cohomology of $\zt $-graded algebras 
was considered by D. Kastler in \cite{8}.
\head \S 4. Definition of $\Ainf $ Algebras \endhead
Suppose that $W$ is a $\zt $-graded space over $k$, and 
denote the tensor algebra over $W$ by 
$ T(W)=\bigoplus_{n=1}^\infty W^n $, where $ W^1=W $ and 
$W^{n+1}=W\tns W^n$. Then $T(W)$ is an associative
algebra under the tensor product.
As usual, the parity of a product 
 $ w=\mtns an$ is given by
$|w|=|a_1|+\cdots+|a_n| $.
\endgraf
\define\hd{\widehat d}
\define\dlk{d_{l,k}}
In this picture, we let $\hd $ be a (super) derivation
on $T(W)$; by this we mean that 
$ |\hd \omega|=|\omega|+|d| $ for homogeneous $ \omega\in T(W)$,
where $|d|$ is the parity of the map $d$,
and also
$\hd (\omega\tns\eta)=
\hd (\omega)\tns\eta +(-1)^{\omega d} \omega\tns \hd \eta $, 
which is the derivation law for (super) derivations.
If we denote the restriction of $\hd $ to $W$ by $d$, then
the derivation law shows that $\hd $ is uniquely
determined by $d$. Denote $d=d_1+d_2+\cdots, $ where
$d_k:W\to W^k$ is the induced map, and similarly denote 
$\hd =\hd_1+\hd_2+\cdots $, where $\hd_l $ is the
component of $\hd $ of degree $l-1$ and where for each $k$,
the restriction $\dlk $ of $\hd_l$ to $W^k$ is the map
$\dlk :W^k\to W^{k+l-1}$ induced by $\hd $. We can express
$\dlk $ in terms of $d_l$ by
$$
\dlk =
\sum_{i+j=k-1} I_i\tns d_l\tns I_j
$$
where $I_i:W^i\to  W^i$ and 
$I_j:W^j\to  W^i$ are the identity maps, with the obvious convention
when either $i$ or $j$ is zero.
It should be noted that if $d$ is an odd (even) map from $W$ to
$T(W)$, 
the maps $d_k$, $\dlk $, and $\hd_k$ are all odd (even) as well.
If $d:W\to  T(W)$ is any map, then it extends uniquely
to a derivation of $T(W)$, so that there is a one-to-one
correspondence between derivations on $T(W)$ and 
maps from $W$ to $T(W)$. 
\endgraf
\define\hdd{\widehat{d}{}^2}
We say that $\hd $
is a differential if $ \hdd =0 $. This 
condition is also determined by the mapping $d$ alone. Since 
$$
\hdd (a\tns b)=\hd(da\tns b+(-1)^{ ad}a\tns db)
=\hd d(a)\tns b+a\tns \hd db,
\tag 13
$$
it is clear that the necessary condition $\hd d=0$ is 
sufficient for $ \hdd =0$ as well. Hence,
$$
\sum_{k+l=n}d_{k,l} d_l=0
\tag 14
$$
for all $n>1$.
This yields an infinite set of relations which are
necessary for $ \hdd =0 $, and these are evidently 
sufficient as well. These relations are simpler than
the more complete set of relations 
$\sum_{k+l=n}\hd_k \hd_l=0 $ for all $ n>1 $. Since $ d_{1,1}=d_1 $,
the first relation yields 
$d_1^2=0$, so
that $d_1$ determines a differential on $W$.
Clearly, $\hdd_1=0 $, which shows that 
$\hd_1$ is itself a differential. 
\endgraf
Now the actual object we want to study is the dual of the
object that we have been considering. More precisely,
let $ V=\Pi (W^*)$, where $\Pi $ denotes the parity reversion of $W$.
(The parity reversion $\Pi W$ of a 
superspace $W$ is given by taking the same
underlying space, but assigning opposite parity to each element.) Then
the map $\pi :W\to \Pi W$ given by the identity mapping is an odd map.
A derivation $\hd $ induces a dual map $\widehat{m}:T(V)\to T(V)$, which is
completely 
determined by the dual $m:T(V)\to  V$ of $d$. 
We omit the details of this construction, but note that the signs
which occur in the formulas below are obtained from this 
dualization of the map, using the exchange rule.
\define \mlk{m_{l,k}}
Similarly,
$d_k$ and $\dlk $ induce dual
maps $m_k:V^k\to  V$ and $\mlk :V^{k+l-1}\to  V^k$.
If $\hd$ is a differential, these maps satisfy the condition
$$
\sum_{k+l=n}m_k\mlk =0.
\tag 15
$$
We can express $\mlk $ in terms of $m_l$ as follows:
$$
\mlk =\sum_{i+j=k-1}(-1)^{i(l+1)+m(k-1)}I_i\tns m_l\tns I_j,
$$ 
where $m=|\hd |$ is the parity of $m$.
Notice that for an odd derivation $\hd $, the associated maps $m_l$ and 
$\mlk $ are odd for odd $l$ and even for even $l$, while for $\hd $ is even the
situation is reversed. 
When $\hd $ is an odd differential, the condition
$\sum_{k+l=n+1}m_k\mlk =0 $
yields the relations
$$
\sum \Sb k+l=n+1 \\ i+j=k-1 \endSb
(-1)^{s_{i,l}}
m_k(v_1,\dots ,v_i, m_l(v_{i+1},\dots ,v_{i+l}),
v_{i+l+1},\dots ,v_n)=0,
\tag 16
$$
where $ s_{i,l}=l(v_1+\dots +v_i)+i(l+1)+n-l $.
This set of relations defines the structure of a strongly
homotopy associative algebra, 
also called an $\Ainf $ algebra, on
$V$. (This structure was introduced by Stasheff \cite{15, 16}.)
The consideration above shows that in the case when $V$ is finite-dimensional,
an $\Ainf $ algebra structure on $V$ can be defined in 
terms of operators acting on the tensor algebra $T(W)$ of
$W=(\Pi V)^*$. Namely, the $\Ainf $ structure on $V$ can be specified
by means of an odd derivation $\hd $ of this algebra satisfying 
$\hdd =0 $.
\footnotemark \nopagebreak \footnotetext{
When $V$ is not finite-dimensional, one can give a similar definition
of an $\Ainf $ algebra by replacing $T(W)$ with its completion, but it is
more convenient to dualize the definition by passing from algebras to
coalgebras. Namely, one should introduce a coalgebra structure on
$ \bigoplus_{k=1}^\infty (\Pi V)^k $, and define an $\Ainf $ algebra by
means of a codifferential on this coalgebra, i.e., a coderivation with 
square zero. 
}
\endgraf
For $n=1$, the relation 
(16) becomes $ m_1^2(v)=0 $, which means that $m_1$ is
a differential on $V$. For $n=2$ the relation becomes
$$
m_1(m_2(a,b))-m_2(m_1(a),b)-(-1)^a m_2(a,m_1(b))=0
\tag 17
$$
which says that $m_1$ is a derivation on the product on $V$ determined 
by $m_2$.
For $n=3$, the relation yields
$$
\aligned
m_2(m_2(a,b),c)&-m_2(a,m_2(b,c))=m_1(m_3(a,b,c))+m_3(m_1(a),b,c)\\
&+(-1)^a m_3(a,m_1(b),c)+(-1)^{a+b}m_3(a,b,m_3(c))
\endaligned
\tag 18
$$
which says that $m_2$ is an associative product up to homotopy, which
explains the name given to such a structure. 
\define\hdl{\widehat \delta}
\head \S 5. Deformations of $\Ainf $ algebras \endhead
Now we consider the case of an infinitesimal deformation
of the operator $\hd $. 
Thus we consider a derivation of the form $\hd+t\hdl $, where $t$
is an infinitesimal parameter whose parity should be opposite to
the parity of $\hdl $, so that the resulting operator remains odd.
This operator is a differential when
$ (\hd +t\hdl)^2=0 $, which is precisely the condition
$\hd \hdl -(-1)^{\hdl}\hdl \hd =0 $, or, in other words, $[\hd,\hdl\} =0$,
where $ [\cdot,\cdot\} $ is the superbracket on the 
superderivation algebra of $T(W)$. We use the fact that
the (super)derivations on $T(W)$ are in one to one correspondence
with $\Hm (W,T(W)) $ to 
introduce a differential on $ \Hm (W,T(W)) $ by 
$ \widehat{D(\delta)}=[\hd,\hdl\} $. 
It is easy to check that $D^2=0$.
If we denote $\widehat{\rho }=D(\hdl )$, then we calculate that
$\rho_n=\sum_{k+l=n+1}d_{k,l} \delta_l-(-1)^{\hdl }\delta_{l,k}d_l.$
\endgraf
Now if $\hdl $ is any derivation on $T(W)$, then by the same 
construction as we used to associate $m$ to $\hd$, we associate
an element $\mu $ of $\Hm (T(V),V)$ to $\hdl $.
We define the parity of the associated map to be the same as 
the derivation it is associated to, but as a map we note that 
$m_l$ has parity $l$, and more generally, the parity of 
$\mu_l$ is $|\delta|+l+1$. If $\widehat\nu $ is the map associated to
$\widehat\rho $, then we compute that  
$$
\nu_n=\sum_{k+l=n+1}m_l\mu_{k,l}-(-1)^{\widehat \mu }\mu_l m_{k,l},
\tag 19
$$
and this process defines a differential $ D(\mu)=\nu $ 
on $ \Hm (T(V),V) $. We use the same notation for the differential
on $ \Hm (T(V),V)$ as on $ \Hm (W,T(W))$.
Thus we can consider the homology groups determined by these
differentials. These homology groups coincide in our finite-dimensional case.
We say that the homology obtained in 
this manner is the Hochshild cohomology of the $\Ainf $ algebra
$V$, and denote it by $HH(V)$.
\endgraf
For convenience in the formulas to follow, we make the following 
sign convention:
$$
s_{i,l,\mu,n}=(l+\mu+1)(v_1+\dots + v_i) +(l+1)i+\mu(n-l),\quad (i\ge 1)
\tag 20
$$
If $\nu =D(\mu ) $, then we have
$$
\aligned
\nu_n&(v_1,\dots , v_n) \\
&=\!\!\!\sum \Sb k+l=n+1 \\ 0\le i\leq k-1 \endSb
(-1)^{s_{i,l,\mu ,n}}
m_k(v_1,\dots ,v_i,\mu_l(v_{i+1},\dots ,v_{i+l}),v_{i+l+1},\dots ,v_n) \\
&\quad-\!\!\!\sum \Sb k+l=n+1 \\ 0\le i\leq l-1 \endSb
(-1)^{s_{i,k,m,n}+\mu}
\mu_l(v_1,\dots ,v_i, m_k(v_{i+1},\dots ,v_{i+k}),v_{i+k+1},\dots ,v_n)
\endaligned
$$
The kernel of this differential is the space of all infinitesimal
deformations of the $\Ainf $ algebra. If $m$ is the
collection of maps $m_k:V^k\to V$, which we call the multiplications
in $V$, and $\mu\in \Hm (T(V),V) $, then $\mu $ determines 
an infinitesimal deformation $m+t\mu $ of $m$ precisely when
$D(\mu )=0$. This interpretation is the same as in the case of an
associative algebra. However, when we consider what a trivial 
deformation is, we note that it no longer is determined by a map
$\rho_t:V\to  V $, as in the case of associative algebras. 
Namely, we define a trivial deformation of an $\Ainf $ algebra by
means of infinitesimal
automorphisms of the tensor algebra $T(W)$, or, equivalently, 
infinitesimal automorphisms of the cotensor algebra 
$\bigoplus_{k=1}^\infty (\Pi V)^k $. Such an automorphism has the
form $\widehat \rho_t =I+t\widehat \lambda $, where $ \widehat \lambda $ is a
derivation 
of $T(W)$, and it is easy to check that the corresponding change of the
differential is given by the formula $\hd \to \hd +tD(\widehat \lambda ) $.
This means that the infinitesimal deformations are classified by the
homology $HH(V)$. 
\head \S 6. Cyclic cohomology of $\Ainf $ algebras \endhead
We give a definition of the cyclic cohomology of an $\Ainf $
algebra and prove that the infinitesimal deformations of an
$\Ainf $ algebra preserving an invariant inner product
are classified  by the cyclic cohomology. 
Suppose that $\tmu \in \Hm (T(V),k)$.
Then we define $\dd (\tmu )$ by
$$
\aligned
\dd(\tmu)_{n+1}(v_1,\dots , v_{n+1})&=\!\!\sum \Sb k+l=n+1 \\ 0\le i\leq  n \endSb
(-1)^{(v_1+\dots +v_{[i+l]})(v_{[i+l+1]}+\dots +v_{n+1})+l(n+1)+ni+\mu}\\
&\quad\times\tmu_l(m_k(v_{[i+l+1]},\dots ,v_{i}),
v_{[i+1]},\dots ,v_{[i+l]})
\endaligned
\tag 21
$$
\endgraf
An element $\tm $ of $\Hm(T(V),k)$ is said to be cyclic if
$$
\tmu_n(v_1,\dots ,v_{n+1})=
(-1)^{n+v_{n+1}(v_1 +\dots +v_n)}\tmu_n(v_{n+1},v_1,\dots ,v_n).
\tag 22
$$
\endgraf
One can check the following fact:
{\it If $\tmu $ is cyclic, then $\widetilde \nu =\dd (\tmu ) $ is also cyclic}.
\endgraf
The fact that $\dd $ is a differential on the cyclic elements of
$ \Hm (T(V),k) $ will follow from the considerations below.
We define the cyclic cohomology $HC (V)$ to be the homology determined
by this differential.
\endgraf
Suppose that an $\Ainf $ algebra $V$ is equipped with an inner product
$\inpr{\cdot }{\cdot } $, and that $ \mu\in \Hm (T(V),k) $ is induced 
by the maps $\mu_k:V^k \to V$.
Then $\inpr{\cdot }{\cdot } $ is said to be invariant with respect to 
$\mu $ if the maps $ \tmu_k:V^{k+1}\to k $, given by 
$$
\tmu_k(\mtns v{k+1})=\inpr{\mu_k(\mtns vk)}{v_{k+1}}
\tag 23
$$
are cyclic.
\endgraf
The inner product induces a map $ \Hm (T(V),V)\to \Hm (T(V),k) $, by
associating the map $\tmu $ to $\mu $, where $\tmu_k$ is given by 
formula (23). When this is an isomorphism, we can use it to define
a differential $\dt $ on $ \Hm (T(V),k) $, by the rule 
$ \dt (\tmu )=\widetilde{D(\mu )}.$
\endgraf
It follows that $\widetilde\nu =\dt (\tmu ) $
is given by the formula
$$
\aligned
&\widetilde \nu (v_1,\dots ,v_{n+1}) \\
&=\sum \Sb k+l=n+1 \\ i\leq k-1 \endSb
(-1)^{s_{i,l,\mu,n}} \tm_k
(v_1,\dots ,v_i,\mu_l(v_{i+1},\dots ,v_{i+l}),v_{i+k+1},\dots ,v_{n+1})\\
&+\sum \Sb k+l=n+1 \\ i\leq l-1 \endSb
(-1)^{s_{i,k,m,n}+\mu +1}
\widetilde \mu_l
(v_1,\dots ,v_i, m_k(v_{i+1},\dots ,v_{i+k}),v_{i+k+1},\dots ,v_{n+1}).
\endaligned
$$
If we assume temporarily that both $m$ and $\mu $ are cyclic with 
respect to the inner product, then we can express the differential as
follows:
$$
\aligned
&\dt (\tmu )(v_1,\dots ,v_{n+1}) \\
&=\sum \Sb k+l=n+1 \\ 0\le i\leq k-1 \endSb
(-1)^{(v_1+\dots + v_{i+l})(v_{i+l+1}+\dots + v_{n+1})+l(n+1)+ni+\mu}\\
&\quad\times\tmu_l
(m_k(v_{i+l+1},\dots ,v_{n+1}, v_1,\dots ,v_i),v_{i+1},\dots ,v_{i+l}) \\
&+\sum \Sb k+l=n+1 \\ 0\le i\leq l-1 \endSb
(-1)^{(v_1+\dots + v_i)(v_{i+1}+\dots + v_{n+1})+ni+l+\mu }\\
&\quad\times\tmu_l
(m_k(v_{i+1},\dots ,v_{i+k}),v_{i+k+1},\dots ,v_{n+1}, v_1,\dots ,v_i)
\endaligned
\tag 24
$$
Now we drop the assumption that $m$ is cyclic and
define a new operator $\dd $ on cyclic
elements of $ \Hm (TV,k) $ by the formula above. This map coincides with
the map defined in the beginning of this section.
It is straightforward to see that $\dd $ is a differential. 
\endgraf
{}From the foregoing, we see that deformations of an $\Ainf $
algebra that preserve an invariant inner product are classified by cyclic 
homology.
Trivial 
deformations are given by 
infinitesimal 
automorphisms of the coalgebra associated to $V$ preserving
the inner product.

\head \S 7. $\Ainf $ deformations of associative algebras \endhead

Suppose that $V$ is actually an associative algebra, so that
$m_2$ is the associative product and all other multiplications
vanish. Then we can consider the deformations of $V$ into an
$\Ainf $ algebra. These deformations are given by the coboundary
operator of $\Ainf $ cohomology. Now suppose that $\mu $ has only
one term $\mu_k$.
Comparing the coboundary operator with the Hochschild coboundary
operator in the $\zt $-graded associative algebra case, one sees that
the Hochschild coboundary coincides with the $\Ainf $ coboundary. 
Therefore, we see that $\Ainf $ deformations
of an associative algebra are actually classified by the Hochshild 
cohomology. In other words, an $\Ainf $ deformation is determined
by an element in $\prod_{k=1}^\infty {HH }^k(V) $. Similarly, one sees
that deformations of an associative algebra with an
(invariant) inner product to an $\Ainf$ 
algebra with an inner product are given by the cyclic cohomology
$\prod_{k=1}^\infty HC ^k(V) $.

\head \S 8. Second Order Deformations \endhead

We show that the cohomology ring $HH (V)$ possesses a natural 
structure of a Lie (super) algebra. To see this, consider the dual
picture again, with $\hd $ being a derivation of the tensor algebra $T(W)$.
We saw that elements of $ \Hm (W,T(W))$ correspond to derivations of $T(W)$.
Thus we have a bracket on $ \Hm (W,T(W))$ given by the bracket of
derivations. Recall that the differential is given in terms of this
bracket by $D(\delta )=[d,\delta \}$. It is easy to check that the bracket
descends to a bracket on the cohomology.
In our finite-dimensional picture, this induces a bracket on the 
cohomology $HH(V)$. In the general situation, this is still true,
but one uses the the bracket of coderivations to  define the bracket
structure. Actually, more is true. It is known (see 
\cite{13, 12}) that the cohomology group of a
differential Lie algebra has the natural structure of an 
$L_\infty $-algebra (strongly homotopy Lie algebra). Applying this 
statement to the situation above we can provide $HH (V)$ with the
structure of an $L_\infty $ algebra. Similar considerations give a
Lie algebra structure (and moreover an $L_\infty $ structure) on the
cyclic cohomology $HC (V)$ of an $\Ainf $ algebra $V$ with an invariant
inner product. 
It is important to stress that this structure depends
on the choice of the inner product.
\endgraf
\define\bardelta{{\overline{\delta}}}
Now we consider a second order deformation of $\hd $. It is given by
$\hd_t=d+t\delta+t^2\epsilon $, where we set $t^3=0 $. We assume here
that the parameter $t$ is even, although there exists a more general 
definition. Then the condition $ d_t^2=0 $ is equivalent to  $ D(\delta)=0$ 
and $ [\delta,\delta]=2D(\epsilon) $. Thus $\delta $ is a cocycle. Denote
its image in homology, by $\bardelta $. Then the second condition
means that $[\bardelta ,\bardelta ]=0$, which is a necessary and
sufficient condition for extending the first order deformation
$ d+t\delta $ to a  second order one. In other words, 
$[\bardelta,\bardelta ]$ is the first 
obstruction to extending the infinitesimal deformation to a formal
deformation of the algebra.

\head \S 9. Ribbon Graphs \endhead

Let us consider a ribbon graph with each vertex having at least three
edges. By definition, a ribbon graph (fatgraph)
is a graph together with a fixed cyclic
order of the edges at each vertex. 
Since all graphs we consider will be ribbon graphs, we will omit the
word ribbon from now on.
We say that a graph is equipped with
a metric if a positive number is assigned to each edge. The set
$\sg $ of all metrics on the ribbon graph can be identified with 
$\R_+^k $, where $k$ is the number of edges in $\Gamma $. (Here 
$\R_+^k$ denotes the subset of $\R^k $ consisting of points with positive
coordinates.) Therefore, $\sg $ is topologically a cell. There is a 
standard construction of a closed surface corresponding to a ribbon graph.
If the graph is equipped with a metric then the surface can be provided
with a complex structure. Let $\Rgn $ be the union of the cells $\sg $, 
where $\Gamma$ corresponds to a surface of genus $g$ with $n$ punctures.
This set has a natural topology. The limit when the length of one of the
edges tends to zero corresponds to the contraction of this edge. 
It is well known \cite{7, 15} that $\Rgn $ is topologically equivalent to 
$ \Mgn \times\R^n_+ $, where $\Mgn $ is the moduli space of compact complex
curves of genus $g$ with $n$ marked points. The decomposition of $\Rgn $
into the cells $\sg $ is not a cell complex; however, one can define the
boundary of a cell $\sg $ in the usual manner  and thus define 
the corresponding homology. This homology is closely related to 
the homology of $\Mgn $.
\endgraf
To define the homology, we need to examine the complex more carefully.
First, we notice that there is an obvious method of deciding when two
graphs are equivalent, determining the same cell. We also need a notion
of an orientation of the cell corresponding to a graph, which is given
by choosing a labeling of the edges and an ordering of the holes in
the graph. Then two oriented cells are equivalent if there is a graph
equivalence between them such that the orientation induced by the 
mapping agrees with the chosen orientation of the cell.
\endgraf
It is known that $\Mgn $ has a natural complex structure,
so that given a choice of the ordering of the holes in the graph,
which determines an orientation of $\R^n_+$, one obtains a canonical
orientation on the highest dimensional cells in $\Rgn $.Note that 
the highest dimensional cells are those corresponding to trivalent 
graphs.
\endgraf
In \cite{10}, there is a general construction 
that 
uses an arbitrary $\Ainf $ algebra to
associate a function to oriented graphs.
Before
considering this general construction, we examine a 
simplified version, which is related to the construction given in
\cite{4}.
\endgraf
Suppose that $V$ is an associative (super) algebra with a nondegenerate 
(super) symmetric bilinear form $h$. Suppose that $\{e_i\}_{i\in I} $ is 
a basis of $V$, and the structure constants $ m^k_{ij} $ are given by 
$ e_ie_j=m^k_{ij}e_k $. We use the matrix $ h_{ij}=h(e_i,e_j) $, and 
its inverse $ h^{ij} $  to raise and lower the indices. The lower 
structure constants $ m_{ijk}=h(e_ie_j,e_k)=m^l_{ij}h_{kl} $
are cyclically (graded) symmetric, so that 
$ m_{ijk}=(-1)^{e_k(e_i+e_j)}m_{kij} $. Similarly, we note that 
$ h^{ij}=(-1)^{e_ie_j} h^{ji} $. We assign a number to a trivalent
graph as follows. To each edge in the graph we associate
two indices, one for each vertex, and the tensor 
$ h^{ij} $, where $i$, $j$ are the indices associated to the edge.
To each vertex we assign the tensor $ m_{ijk} $, where $i$, $j$, $k$ 
are the indices associated to the incident edges, and the order 
is chosen to be consistent with the cyclic order of the edges at the vertex.
We multiply all of these symbols together and sums over repeating 
indices to obtain a number $Z(\Gamma )$.\footnotemark
\nopagebreak \footnotetext{In the graded case, one also picks up a sign 
in each term depending on the parity of the basis elements, since what we 
are really doing here is performing a series of graded contractions of 
a tensor according to the prescription which is dictated by the graph.}
\endgraf
The resulting function $Z(\Gamma)$ depends only on the
genus $g$ and the number $n$ of holes in the graph. This is easy to
see, since every trivalent graph with the same number of holes and
the same genus can be obtained from one such graph by performing 
a series of simple transformation, called the fusion move
which is illustrated below.
\midinsert
\HideFigureFrames
\BoxedEPSF{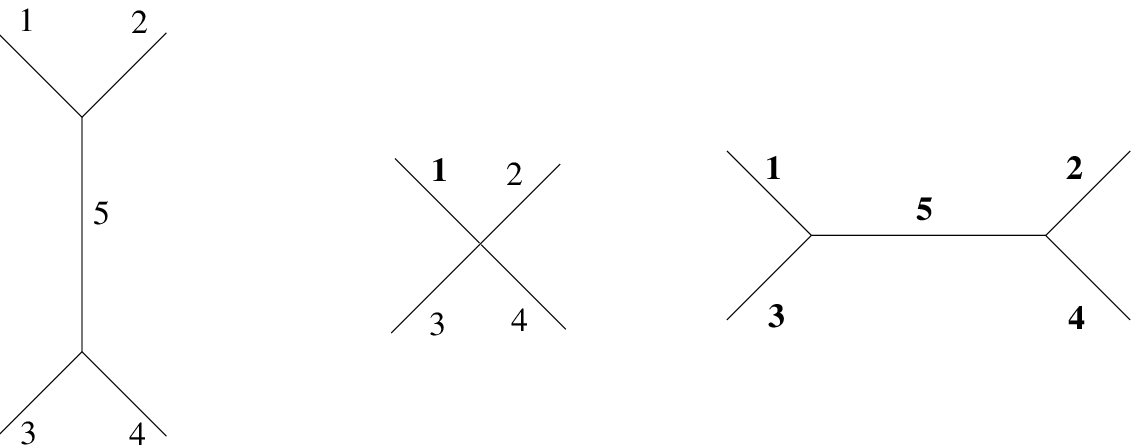}
\botcaption{Figure 1} Fusion diagram 
\endcaption
\endinsert

It is easy to calculate how 
the functions of two graphs differing by a fusion move are 
related, and the associativity of the algebra guarantees that 
the functions will have the same value.
Thus if $V$ is equipped with the structure of an associative
algebra, we can use the structure constants 
$ m_{ijk}= \tm _2(e_i,e_j,e_k) $ to obtain a 
function that depends only on the genus and number of holes in the 
graph. We note that in writing down the function $Z(\Gamma)$,
an order of the vertices and an order of the edges must be chosen,
but the resulting function is independent of these orders because
the tensors $m_{ijk}$ and $h^{ij}$ are even, so that the corresponding
contractions are the same. We also see that the form of the 
function $Z(\Gamma)$
depends on the starting edge at each vertex, which determines
the order in which the indices for $m_{ijk}$ are listed, but 
the cyclic symmetry of this tensor again ensures that the function
is independent of this choice. Finally, the same arguments show that 
the function is independent of the order in which the tensor 
$ h^{ij} $ is presented, due to the symmetry of the inner product.
\endgraf
We would like to construct homology classes of the complex $\Rgn $.
With this goal in mind, we construct chains in this complex, that is, 
linear combinations of oriented cells. Since a change of the orientation 
corresponds to a change of sign in the complex, we want to define 
a function on oriented cells that  changes sign when the orientation 
of the cell is reversed. For simplicity, we identify cells with graphs, 
and therefore replace oriented cells with oriented graphs. Note that this
is not the usual notion of an oriented graph.
In the discussion above, the function $Z(\Gamma)$ was defined on 
the set of trivalent graphs. What we really want is to define a
function on the set of oriented graphs $\Gor $, which
we shall denote $Z(\Gor )$. 
For an oriented trivalent graph $\Gor $, we define $Z(\Gor )=Z(\Gamma)$
if the orientation is the canonical one, and $Z(\Gor )=-Z(\Gamma)$
otherwise.
It is easy to see that the element 
$$
Z_{\text{max}}=\sum Z(\Gor )\Gor ,
\tag 25
$$
where the sum is taken over all oriented trivalent graphs, is a cycle in
the homology of $\Rgn $ discussed above, since $Z(\Gor )$ is constant
on all graphs with the canonical orientation. This result is just 
the assertion that the cell complex is orientable. Of course, the orientation
in this simple case is somewhat superfluous, but in the more general
construction to follow, the orientation is quite relevant.
\endgraf
Now suppose that $\varphi_k$ is a cyclic $k$-cocycle on the algebra $V$. 
We construct a cycle of dimension $k-2$ less than the maximal dimension 
on $\Rgn $. Consider graphs in which all vertices except one are trivalent, 
and the exceptional vertex has $k+1$ incident edges. As before, we
assign to each trivalent vertex the tensor $ m_{ijk} $, and to the
exceptional edge we assign the tensor 
$\varphi_{i_1\cdots i_{k+1}}=\widetilde\varphi(e_{i_1},\dots ,e_{i_{k+1}})$.
One constructs $Z(\Gamma )$ in the same manner as before, 
but now there is a problem in the definition if $k$ is odd, because 
in this case the cyclicity of $\widetilde\varphi_k$
means that the formula for $Z(\Gamma)$ depends
on which edge one starts with. Of course, the function
is determined up to the total sign, since in the expression
$$
\varphi_{{i_1}\cdots i_{k+1}}=(-1)^{k+e_{i_{k+1}}(e_{i_1}+\dots +e_{i_k)}}
\varphi_{i_{n+1}i_1\cdots i_k}
\tag 26
$$
the sign $ (-1)^{e_{i_{k+1}}(e_{i_1}+\dots + e_{i_k})} $ 
cancels because this is a graded contraction.
Thus only the $ (-1)^k $ plays a role. 
Therefore, in order to assign a fixed value we must consider 
a starting edge for the vertex.
\endgraf
To explain this, we introduce the notion of a ciliated graph
$\Gc $, which is a ribbon graph with a preferred edge chosen for
each vertex. This terminology was suggested by Fock and Rosly,
see \cite{3}.
 Given a ciliated graph with at most one non trivalent
edge, one obtains a well-defined formula for $Z(\Gc )$ by using
the preferred edge to determine the order in which to write down
the terms for the vertices. Of course, for the trivalent vertices,
the choice of the cilia does not affect on the outcome, but for the
exceptional vertex the choice is relevant if $k$ is odd. Now we 
also note that in this case the order of the vertices is not
important, because there is at most one vertex corresponding to
an odd tensor, so this case is still independent of the order.
\endgraf
We want to define a partition function on oriented graphs that
takes opposite signs on graphs of opposite parity. To do
this we also use the canonical orientations of trivalent 
graphs as follows. For the exceptional vertex, 
we can expand the vertex by inserting edges, using the ciliation as
a starting point, to obtain a trivalent graph, as illustrated in the
figure below.
\midinsert
\HideFigureFrames
\BoxedEPSF{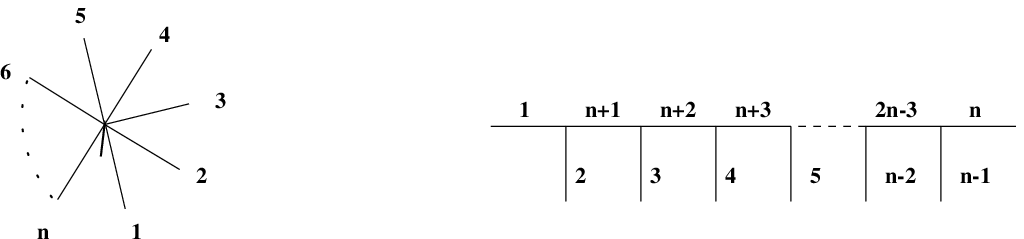}
\botcaption{Figure 2} Expanding a ciliated vertex
\endcaption
\endinsert
An orientation of the original graph induces an orientation on the
trivalent graph, which we can compare to the canonical orientation.
If the number of incident edges to the vertex is even, the comparison
will depend upon where the cilia is placed, and will alternate as one
moves the cilia one edge at a time. On the other hand, if the number
of edges is odd, then the orientation of the expanded graph will not
depend on the placement of the cilia.
Thus we can define the 
function $Z(\Gor )=\pm Z(\Gc )$, where the sign is plus if the expanded
graph associated to $\Gor $ with the given ciliation has the canonical
orientation. Note that result does not depend on the ciliation.
One can prove that the element
$$
Z=\sum Z(\Gor )\Gor 
\tag 27
$$
where the sum runs over all distinct\footnotemark
\nopagebreak\footnotetext{
Note that in this construction a graph with opposite orientation is
the negative of the oriented graph, so is not considered as distinct
in our consideration.}
oriented graphs with exactly one nontrivalent vertex with $k+1$ incident
edges, is a cycle. We omit a direct proof of this statement. Instead we 
will derive this result from the fact that a cyclic cocycle determines 
an infinitesimal deformation of an associative algebra 
with an invariant inner product into an $\Ainf $ algebra 
with an invariant inner product. Applying a
general result of Kontsevich which we shall explore below, one sees
immediately that $Z$ is a cycle.
\endgraf
Now let us consider the case when $V$ is an $\Ainf $ algebra, and 
graphs with only the restriction that each vertex has at
least three edges. We can repeat the construction of the
function $Z(\Gamma)$ as before, associating to each vertex with
$k+1$ edges the tensor $\tm_{i_1\cdots i_{k+1}}$. The result will 
depend on the order in which the
tensors corresponding to the vertices are listed, more precisely,
the order in which the vertices with an even number of incident
edges are listed. However, we can resolve this in a similar way
by considering graphs with a fixed ciliation and a fixed order of vertices,
for which the partition function is well defined, and then multiplying
by a sign that depends on the order, ciliation, and
orientation, in such a manner that the resulting partition function
depends on the orientation alone. To do this, we expand those vertices
that are not trivalent,  as we did before, but now we
must also keep track of the order 
of the vertices when adding the new labels, so
that we not only get an orientation of the graph, but an order of
the new holes as well. The canonical orientation is determined by
the canonical ordering of the holes as well, so that interchanging the
order of the expansion for two vertices will actually produce a
reversal of the sign precisely when both vertices are even.
But this reversal of sign is mirrored in the contraction as well,
so the effect cancels. The ciliation effect is treated as before,
so that again we are able to define a function $\Z(\Gor )$
depending only on the orientation of the graph.
\endgraf
A result formulated by Kontsevich in this case is that the chain
$$
Z=\sum Z(\Gor )\sigma_{\Gor }
\tag 28
$$
is a cycle. To see why this is true, note that it is sufficient to
restrict the sum to graphs with a fixed number of edges, and to show
that this partial sum is a cycle. Taking a graph with one edge less
than this fixed number, we consider the ways to expand this graph
by adding one edge. We can consider the graphs obtained by expanding
each nontrivalent vertex separately. Restricting attention to
those obtained by expanding a single vertex by inserting an edge in
one of the possible manners, we obtain expressions for
the function that differ only in the contribution from the
vertex. If the vertex has $n$ edges, then we will obtain expressions
of the form $ m_{i_1\dots i_{l}}m_{i_{l+1}\dots i_{n+2}} $,
multiplied by a sign that is determined by the expansion into a
trivalent graph. We will show that the sign coincides with the
signs given in the expressions for the relations satisfied by the
$\Ainf $ algebra.
\endgraf

The defining relations for an $\Ainf$ algebra with
an invariant inner product can be stated in the form
$\dd(\tm)=0$, from which, using equation 21, we obtain
$$
\multline
\sum_{{k+l=n+1}\atop{0\le i<=n}}
(-1)^{(v_1+\cdots+ v_{[i+l]})(v_{[i+l+1]}+\cdots+ v_{n+1})+l(n+1)+ni+
\mu}\\
\times\tm_l(m_k(v_{[i+l+1]},\dots, v_{i}),
v_{[i+1]},\dots, v_{[i+l]})=0
\endmultline
\tag 29
$$
These relations yield the expression
$$
\sum_{{k+l=n+1}\atop{0\le i<=n}}{-1}^s
m_{a,j_{[i+1]},\dots, j_{[i+l]}}m_{j_{[i+l+1]},\dots, j_i,b}=0
\tag 30
$$
where
$$s=(e_1+\cdots+ e_{[i+l]})(e_{[i+l+1]}+\cdots+ e_{n+1})+ni+l+1.$$
We use these relations to explain the Kontsevich result. We shall
concern ourselves here with the factor $ni+l+1$ which appears in the
exponent, as the other part of this exponent is cancelled because
this is a graded contraction.

Let us consider a fixed graph, and consider all possible ways
to expand this graph into a graph with one additional edge inserted at
a fixed vertex. Suppose that this fixed vertex has $n+1$ edges, where
of course, $n\ge 3$.  The insertion of an edge will create a graph
with two new vertices, having $l+1$ and $k+1$ edges, where $k+l=n+1$.
Let us suppose that the original vertex had incident edges $1, \dots, n+1$
with
associated labels $j_1$, \dots, $j_{n+1}$. We insert an edge such a
manner that the  new graph will
will have a vertex with labels $a$, $j_{[i+1]}$, \dots, $j_{[i+l]}$,
and one with
labels $j_{[i+l+1]}$, \dots, $j_{i}$, $b$,
where $a$ and $b$ are the indices attached to the new edge, which has
label $n+2$.
The figure 
below shows illustrates the labeling of the expanded graph.
\midinsert
\HideFigureFrames
\BoxedEPSF{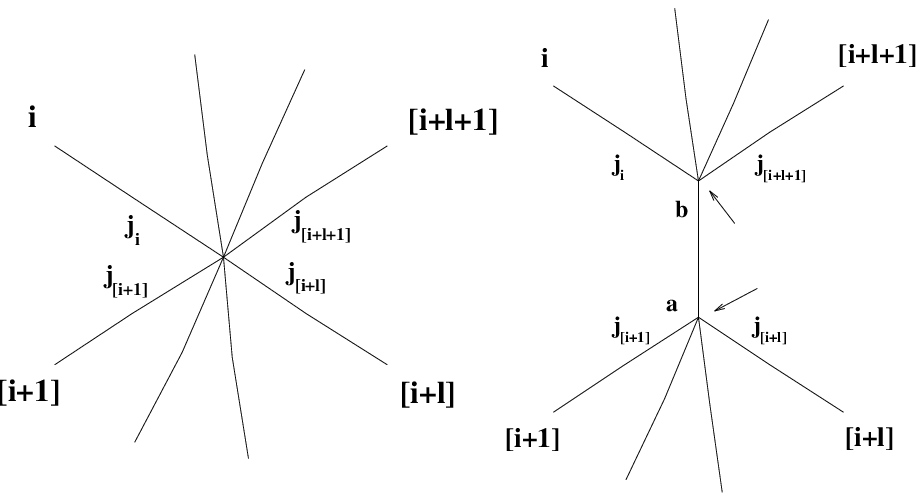}
\botcaption{Figure 3} Splitting a vertex
\endcaption
\endinsert
Inserting cilia
between $b$ and  $j_{[i+l+1]}$, and
 $j_{[i+l]}$ and $a$  respectively, the contribution
of the two vertices to the  function $Z(\Gc)$ is
$
m_{a,j_{[i+1]},\dots, j_{[i+l]}}
m_{j_{[i+l+1]},\dots, j_{[i]},b}
$.
When expanding
the graph according to this ciliation, one sees that the expanded graph
is identical to the one which would be obtained from the original graph,
with a cilia placed between $j_{[i+l]}$ and $j_{[i+l+1]}$. 
However, the orientation
of the expanded graph is affected by the fact that the inserted edge has
been labeled first, instead of in the order in which it would have been
labeled by the usual expansion. In case the original vertex had an odd
number of edges, this is the only factor affecting the orientation,
because the placement of the cilia in the original vertex is irrelevant.
Since $n$ is even, the factor $ni$ does not contribute in the sign
$ni+l+1$.
Thus one picks up a factor which depends only on $l$. On the other hand,
when the original vertex had an even number of edges, the factor $ni$
does contribute, which reflects the fact that the placement of the
cilia is relevant to the sign of the function $Z(\Gc)$.
When we sum over all graphs which result from
the insertion of an edge at the fixed vertex, these considerations and
equation 30 show that the sum is zero, and this is precisely
the incidence number of the graph in the boundary of the chain $Z$. This
shows that this chain is a cycle.
\endgraf
We have tacitly assumed that algebra in question does not have any
product of degree one, as the terms
involving such a product are not accounted for. However, there are some
cases when one can include such a multiplication. We shall not go into this
matter here.
\endgraf
The construction of a cycle in $\Rgn $ by a cyclic cocycle of an 
associative algebra can be considered as a limiting case of 
Kontsevich's construction. We can consider this as an infinitesimal
deformation of the algebra into an $\Ainf $ algebra. On the other
hand, the proof can be carried out directly because the signs in
equation 12  coincide with those in equation 30.
We can
also apply second order deformations in the context above to construct
other interesting examples of cocycles.
\endgraf
We wish to thank Dmitry Fuchs and Jim Stasheff,
who read this manuscript and made helpful
suggestions. We also would like to express our gratitude to 
Ezra Getzler and Maxim Kontsevich for useful conversations.
A. Sch. is grateful to the MSRI and the I. Newton Institute
for Mathematical Sciences for their hospitality.

\enddocument